\documentclass{emulateapj}
\usepackage{epsf}
\usepackage{apjfonts} 
\usepackage{xspace} 
\usepackage{amsmath}
\def\HI{{\rm H\,I}}
\def\HII{{\rm H\,II}}

\def\Msun{\, M_{\odot}}

\def\fesc{f_{\rm esc}}

\def\Luv{L_{\rm UV}}
\def\Mtot{M_{\rm TOT}}
\def\smin{s_{\rm min}}

\def\dim#1{\mbox{\,#1}}
\def\figname#1#2{#2}
\def\note#1{}
\def\hide#1{}

\begin{document}

\title{Are There Enough Ionizing Photons to Reionize the Universe by $z\approx6$?}

\author{Nickolay Y.\ Gnedin\altaffilmark{1,2,3}}
\altaffiltext{1}{Particle Astrophysics Center, Fermi National
Accelerator Laboratory, Batavia, IL 60510, USA; gnedin@fnal.gov}
\altaffiltext{2}{Kavli Institute for Cosmological Physics, The
University of Chicago, Chicago, IL 60637, USA}
\altaffiltext{3}{Department of Astronomy \& Astrophysics, The
University of Chicago, Chicago, IL 60637 USA}

\begin{abstract}
An estimate for the number of ionizing photons per baryon as a
function of redshift is computed based on the plausible extrapolation
of the observed galaxy UV luminosity function and the latest results
on the properties of the escape fraction of ionizing radiation. It is
found that, if the escape fraction for low mass galaxies
($\Mtot\la10^{11}\Msun$) is assumed to be negligibly small, as
indicated by numerical simulations, then there are not enough ionizing
photons to reionize the universe by $z=6$ for the cosmology favored by
the WMAP 3rd year results, while the WMAP 1st year cosmology is
marginally consistent with the reionization requirement. The escape
fraction as a function of galaxy mass would have to be constant to
within a factor of two for the whole mass range of galaxies for
reionization to be possible within the WMAP 3rd year cosmology.
\end{abstract}


\section{Introduction}
\label{sec:intro}

Studies of cosmic reionization - especially theoretical ones - have
never been considered as a ``photon-starved'' field. Theorists always
felt free to adjust the emissivity of the ionizing sources, usually
quantified by the escape fraction of ionizing radiation, to adjust the
reionization redshift to their choosing.

This approach was, indeed, justified in the earlier studies, since
until recently limited knowledge existed on the reasonable values for
the escape fraction at high redshifts. However, the latest
observational \citep{gcdf02,flc03,sspa06,cpg07} and numerical
\citep{rs06,rs07,gkc07} studies finally begin to converge on the
values and evolution of the escape fraction of ionizing radiation and
on the relative escape fraction between the far UV and Lyman
limit. Three properties of the escape fraction are particularly
important for reionization studies: (1) the value of escape fraction is
small (a few percent at most), which is an order of magnitude
smaller than is assumed in some reionization modeling, (2) it is weakly
dependent on the 
galaxy mass or star formation rate for large galaxies, and (3) it drops
prodigiously for dwarf galaxies. The first two properties are
reproduced in all recent studies, both observational and theoretical,
and, therefore, are rather robust. The last feature of the escape
fraction has only been seen in simulations of \citet{gkc07} and is
indicated by measurements of \citet{flc03}\footnote{However, the
  characteristic galaxy masses below which the escape fraction drops
  down are somewhat different between those two studies.}, because  
other simulations and observational studies do not yet have either
numerical resolution or sensitivity to resolve dwarf galaxies. 

Another important observational advance that places the study
of reionization on a much more quantitative footing is the
observational determination of the galaxy UV luminosity function down
to well below $L_*$ at $z\ga6$. Since it is not possible to give a
comprehensive review of all observations in this {\it Letter\/} due to
space limitations, I refer the reader to the recent work by
\citet{biff07}, who give a detailed review of the current status of
existing observational data. While the data for the galaxy luminosity
function during the reionization era ($z>6$) are still sparse, the
plausible extrapolation of the observed $z\approx6$ luminosity
functions to higher redshifts can be used to predict the global
production of the ionizing radiation to at least within a factor of 2
to 3, i.e.\ more than an order of magnitude improvement over the
previous, purely theoretical, assumptions.

Of course, the most accurate models are only possible with the
large-box and high-resolution cosmological simulations, which model in
detail the emission of ionizing photons in high redshift galaxies and
quasars, the propagation of ionizing radiation in the expanding
universe, and absorption of that radiation at cosmic 
ionization fronts and Lyman limit systems. But even the simplest
balance of the available ionizing photons and the number of atoms that
need to be ionized before the end of reionization at $z\approx6$ - as
required by the observed transmitted flux in the spectra of high
redshift quasars discovered by the SDSS collaboration \citep{fsbw06} -
is a useful exercise after the recent improvements in our understanding
of the sources of reionization. 

This is the subject of this {\it Letter\/}.

\section{Results}
\label{sec:results}

In order to compute the total number of ionizing photons available for
reionizing the universe at any given redshift, the observed luminosity
functions need to be extrapolated to earlier redshifts. Such
extrapolation is, of course, not unique. However, since the mass
function of dark mater halos can be computed sufficiently precisely in
a given cosmology at any redshift, the extrapolation of the luminosity
function to $z>6$ can be made reliably if the relationship between the
galaxy luminosity and the mass of its dark matter halo can be
established.

While such a relationship is unlikely to be a simple function, models that
assume a one-to-one correspondence between the galaxy luminosity and
the halo mass provide remarkably good fit to a variety of
observational tests \citep{cwk06}. Thus, as a simple and crude
approximation, it is instructive to assume such a relationship
between the galaxy UV luminosity $L_{UV}$ for
the high redshift galaxies as well.

\begin{figure}[t]
\plotone{\figname{figFIT.ps}{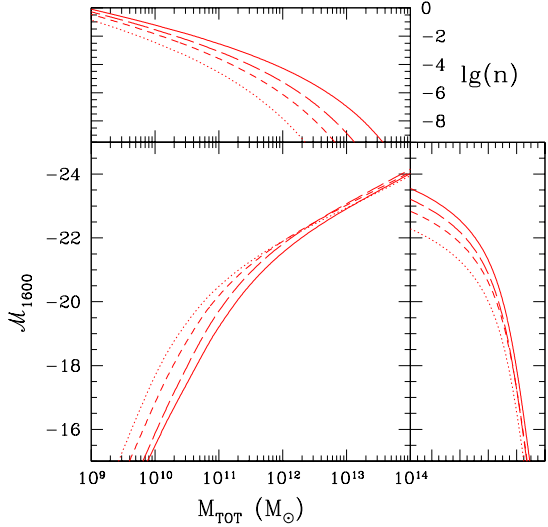}}
\caption{The cumulative mass function of dark matter halos (top
  panel), the cumulative UV luminosity function of galaxies (right
  panel), and the relationship between the UV absolute magnitude and
  the halo mass (central panel) obtained by matching the two. Four
  different lines correspond to $z=3.8$ (solid lines), $z=5.0$
  (long-dashed lines), $z=5.9$ (short-dashed lines) and $z=7.4$
  (dotted lines). The observational data are represented by their
  respective Schechter function fits from \citet{biff07}.}
\label{figFIT}
\end{figure}

\begin{figure}[t]
\plotone{\figname{figM2L.ps}{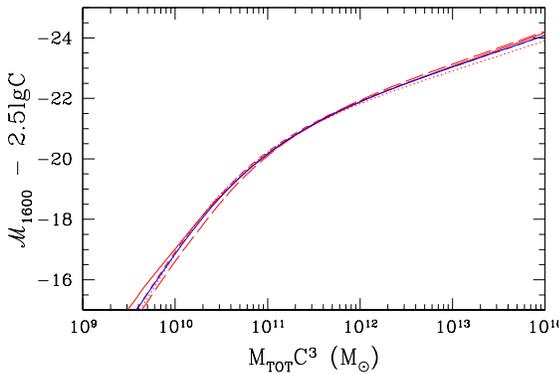}}
\caption{The relationship between the UV absolute magnitude and
  the halo mass from Fig.\ \ref{figFIT}, shifted horizontally and
  vertically using the correction factor $C=(1+z)/7$ (line styles are
  as in Fig.\ \ref{figFIT}). Notice that the
  shift results in a good agreement between different redshifts for
  all galaxies more massive than about $3\times10^{10}\Msun$. The
  blue/black solid line marks the average mass-to-light relationship
  used in the rest of this paper.}
\label{figMTL}
\end{figure}

In a given cosmology, this relationship between the total mass of a
dark matter halo $\Mtot$ and the luminosity of a galaxy hosted in that
halo $\Luv$ can be obtained by matching the cumulative mass and
luminosity functions,
\begin{equation}
  n(>\Luv) = n(>\Mtot),
  \label{eq:nn}
\end{equation}
as is shown in Figure \ref{figFIT} for the 4 values of redshift for
which \citet{biff07} give the parameters of the Schechter function
fits, assuming the best-fit cosmology for the combination of WMAP 3rd
year data and Large Red Galaxies part of SDSS survey
\citep[WMAP3;][]{sbdn06}\footnote{$\Omega_M=0.27$, $h=0.71$,
  $n_S=0.95$, $\sigma_8=0.78$}. Unfortunately, the derived relation
between the UV luminosity (as expressed by the AB absolute magnitude
at $1600\AA$, ${\cal M}_{1600}$) and the total halo mass is
redshift-dependent, and so 
cannot be easily extrapolated to higher redshifts. However, a simple
correction of the UV luminosity by a factor of $C$ and the total mass
by a factor of $C^3$ with 
$$
  C(z) = \frac{1+z}{7}
$$
eliminates most of the redshift dependence for halos more massive than
about $3\times10^{10}\Msun$, as is shown in Figure \ref{figMTL}. As I
discuss below, these low mass halos 
contribute almost nothing to the ionizing photon budget, and are so
unimportant for the purpose of this paper. Throughout the rest of this
paper, I use the average relation marked by a blue/black line in
Fig.\ \ref{figMTL}.

The relation between the UV and ionizing luminosities is quantified by
the {\it relative\/} escape fraction, $f_{\rm esc, rel} =
\fesc(912\AA)/\fesc(1600\AA)$ and the value of the intrinsic ratio of
stellar luminosities at these two wavelength, $r_{\rm int} =
(L_{1600}/L_{912})_{\rm int}$. For the escape fraction at the Lyman
limit I adopt the results of \citet{gkc07}, who found that in high
resolution simulation of galaxies with radiative transfer escape
fractions for larger galaxies are of the order of a few percent,
consistent with observational determinations, but little (if any)
radiation escapes from small galaxies. Thus, for the relative escape
fraction as a function of galaxy mass I adopt the following form:
$$
  f_{\rm esc, rel}(M) \approx 0.15
  \begin{cases}
    1 & \text{if $\Mtot>5\times10^{10}\Msun$,} \\
    \smin & \text{otherwise.}
  \end{cases}
$$
This form is consistent with observational measurements of the
relative escape fraction \citep[c.f.][]{sspa06} for massive
galaxies. The drop in the escape fraction is also indicated in
observations of \citet{flc03}, but the characteristic transition
occurs at a factor of 10 higher star formation rate, which would
correspond to a higher characteristic mass.  The adopted value of
$5\times10^{10}\Msun$ therefore likely biases the estimate production
of ionizing photons up. Here, as well as in the rest of the paper, all
uncertain quantities are chosen so that to insure that my estimate for
the total number of ionizing photons is likely to be an overestimate,
rather than an underestimate. I return to the uncertainty of the main
result in the Discussion section.

In \citet{gkc07} simulations, the relative escape fraction
of low mass galaxies ($\Mtot<5\times10^{10}\Msun$) never exceeds about
0.01, and is often much lower, so I adopt $\smin=0.05$ as my fiducial
value. I consider the effect of parameter $\smin$ in the Discussion
section.

Using the mass-to-light matching from Fig.\ \ref{figMTL}, the mass
dependence can also be recast as the luminosity dependence.

For the intrinsic break, I adopt a value of \citep{sspa06}
$$
  r_{\rm int} \approx 3.
$$
However, \citet{stcf07} argue for a larger value for the intrinsic
break, $r_{\rm int} \approx 6$, and that larger value is also
consistent with Starburst99 spectral synthesis models 
\citep{lsgd99}. Here I again adopt a lower value as a fiducial number
so that not to underestimate the total number of ionizing photons.

With the above assumption, the total emission rate density of the
ionizing photons at a given redshift $z$ can now be expressed as
\begin{equation}
  \dot{n}_\gamma = \int dL_{1600} \frac{f_{\rm esc,
  rel}}{r_{\rm int}} \frac{L_{1600}}{\langle E\rangle}
  \frac{dn}{dL_{1600}}, 
  \label{eq:ng}
\end{equation}
where $\langle E\rangle$ is the average energy of a photo-ionizing
photon (which I take to be $22\dim{eV}$, consistent with typical
spectra of star-bursts \citep{lsgd99}) and $dn/dL_{1600}$ is the comoving UV 
luminosity function obtained from the halo total mass function using
the mass-to-light ratio from Fig.\ \ref{figMTL}. The total number of
ionizing photons per baryon at time $t$ is then
$$
  N_{\gamma/b}(t) = \int_0^t \frac{\dot{n}_\gamma}{n_b} dt,
$$
where $n_b\approx2.5\times10^{-7}\dim{cm}^{-3}$ is the comoving number
density of baryons.

An additional complication in this estimate, however, is introduced by
a possible contribution to the ionizing background from high-redshift
quasars. Unfortunately, this contribution cannot be estimated with the
same method as the contribution from galaxies, because quasars are
known to be short-lived, and halos of a given total mass may or may
not host a quasar at any given time. 

Fortunately, the quasar luminosity function is reasonably well known
all the way to $z\approx5$ \citep{hrh07}. Using the fitting code provided by
\citet{hrh07}, the emission rate density of ionizing radiation from
quasars can be estimated at a range of redshifts $4<z<6$. Such an
estimate agrees remarkably well with an earlier estimate by
\citet{mhr99}: for example, \citet{hrh07} fit at $z=5$ results in
$\log(\dot{n}_\gamma)\approx50.5$, while \citet{mhr99} estimate at
that redshift is $\log(\dot{n}_\gamma)\approx50.6$. At $z=4$ both
estimates give the same value of $\log(\dot{n}_\gamma)\approx50.8$. 

The extrapolation of the \citet{hrh07} luminosity function to $z>5$ is, of
course, highly uncertain. In order to approximately account for
possible uncertainties, I consider two different extrapolations: the
``lower'' one simply uses the \citet{hrh07} best-fit model to compute
the quasar luminosity function at any redshift; the ``higher''
extrapolation multiplies the ``lower'' one by a factor of
$[(1+z)/6]^3$ (chosen somewhat arbitrarily), greatly increasing quasar
abundance at higher redshifts. 

\begin{figure}
\plotone{\figname{figNGB.ps}{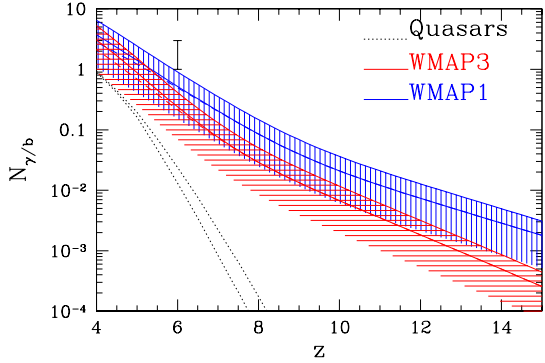}}
\caption{The total ionizing photon-to-baryon ratio as a function
  of redshift for the WMAP3 (red/gray lines) and WMAP1
  (blue/black lines) cosmologies. In both cases the dashed lines show
  the contribution from galaxies only, while solid lines include the 
  quasar contribution (with $\smin=0.05$). The shaded region is
  an estimate of the upper 95\% CL uncertainty (the lower uncertainty
  is harder to estimate, and so its exact limit is not shown). Two
  black dotted lines mark the two estimates of the 
  quasar contribution. The adopted reionization criterion,
  $1<N_{\gamma/b}<3$ at $z=6$, is shown as a segment with
  error-bars.\newline 
} 
\label{figNGB}
\end{figure}

The resultant photon-to-baryon ratio is shown in Figure \ref{figNGB}
for two adopted sets of cosmological parameters: the WMAP3 cosmology
introduced above and the best-fit values for the pure
$\Lambda$CDM model from the first year WMAP data
\citep[WMAP1;][]{svp03}\footnote{$\Omega_M=0.27$, $h=0.72$,
  $n_S=0.99$, $\sigma_8=0.90$}. As can be seen, the quasar
contribution is 
smaller than the one from galaxies by at least an order of magnitude,
and so its large uncertainty is not that important.

The uncertainty on the estimates from Fig.\ \ref{figNGB} is not easy
to evaluate. The observational $2\sigma$ limit on the adopted value of
$f_{\rm esc, rel} \approx 0.15$ is about 50\% \citep{sspa06,cpg07}. In
addition, the model for the evolution of the ionizing luminosity from
galaxies presented above is only approximate. Since the observed
luminosity functions are only used to compute the total number of
ionizing photons (an integral quantity), the observational $2\sigma$ error from
the uncertainty in the luminosity functions is about 40\%
\citep{biff07}. I increase it to 50\% because additional interpolation
is involved, and adding the two uncertainties in quadrature, obtain
about 70\% uncertainty in the upper direction. The lower uncertainty
is much larger, since the selected values for all relevant factors are
consistently biased toward a larger value of $N_{\gamma/b}$. For
example, the value for the intrinsic break $r_{\rm int}$ may be close
to 6 than to 3, resulting in additional factor of 2 uncertainty in the
downward direction. The lower shaded regions in Fig.\ \ref{figNGB}
include this factor-of-2 uncertainty, but the lower bound of the
uncertainty is likely to be even lower, so it is not shown in Fig.\
\ref{figNGB}. 

Thus, it is unlikely that curves in Fig.\ \ref{figNGB} {\it
  underestimate\/} the correct value by more than a factor of two,
  although they can present a substantial overestimate of the correct result.

\section{Discussion}
\label{sec:discussion}

But what is the reionization criterion in terms of
$N_{\gamma/b}$? Obviously, the whole process of reionization
cannot be supplanted by a single value for $N_{\gamma/b}$. On
the other hand, having $N_{\gamma/b}\ga0.9$ is a {\it necessary condition\/}
for reionization (assuming that helium is singly ionized and the mass
fraction of the neutral hydrogen at $z\approx6$ is small, consistent with the
expected abundance of Lyman limit systems where most of neutral
hydrogen resides after reionization). 

On the other hand, a ratio of the the Hubble time to the average
recombination time in a particular region of the universe is 
\begin{equation}
  \frac{t_H}{t_{\rm REC}} \approx 1.3 \left(\frac{1+z}{7}\right)^{3/2}
  \left(1+\langle\delta\rangle\right) C_R,
  \label{eq:thtrec}
\end{equation}
where $\langle\delta\rangle$ is the average overdensity in that
region, and $C_R$ is the recombination clumping factor in that region
\citep{kgh07}, $C_{R}={\langle R(T) n_{e} n_{\HII} \rangle}/({\langle
  R(T) \rangle \langle n_{e}\rangle \langle n_{\HII}\rangle}).$ 

Since the definition of the escape fraction of ionizing radiation from
\citet{gkc07} accounts for all local absorption, including high
density gas inside a galaxy halo, which dominates the clumping factor,
$C_R$ cannot be large in the general IGM. More than that, the
conclusion that reionization is complete by $z\approx6$ comes
primarily from the observations of the SDSS quasars \citep[][and
references therein]{fsbw06}. The transmitted flux at $z\ga5.5$ comes
mostly from the centers of large voids, where
$(1+\langle\delta\rangle)\approx0.1$ and $C_R\la10$. Equation
\ref{eq:thtrec} then implies that less than 2 ionizing photons per
baryon (outside the virial radii of ionizing sources) are needed to
satisfy the observational requirements. Of course, if the local
absorption inside the virial radius (characterized by a large clumping
factor $C_R$) is included, the required number of ionizing photons
per baryon will be much higher; but then a correspondingly larger
value for the escape fraction (which excludes local absorption) should
be adopted.

That estimate is also consistent with the conclusion by \citet{m03},
who estimated that 
$$
\frac{1}{H} \frac{dN_{\gamma/b}}{dt}\la7
$$
for $6<z<9$, which translates into $N_{\gamma/b}\la2.5$ if the
contribution of sources beyond $z=9$ is unimportant.

Thus, a requirement
\begin{equation}
  1<N_{\gamma/b}(z=6)<3
  \label{eq:crit}
\end{equation}
appears to be a sensible criterion for the reionization of the
universe by $z=6$. The same condition has also been obtained by
\citet{bh07} from extrapolating the production rate of
ionizing photons required to fit the observed evolution of the mean
opacity of the Lyman-alpha forest to $z=6$.

Thus, the WMAP1 cosmology is marginally sufficient to satisfy the condition
(\ref{eq:crit}), while the WMAP3 universe is well short of the needed
amount of ionizing radiation at $z\approx6$ by at least a factor of 2
(and, perhaps, as much as a factor of 10 if the value for the intrinsic
break $r_{\rm int}$ is close to 6 than to 3, and the transition to low
escape fraction occurs at $5\times10^{11}\Msun$ rather than at
$5\times10^{10}\Msun$).  

\begin{figure}
\plotone{\figname{figMIN.ps}{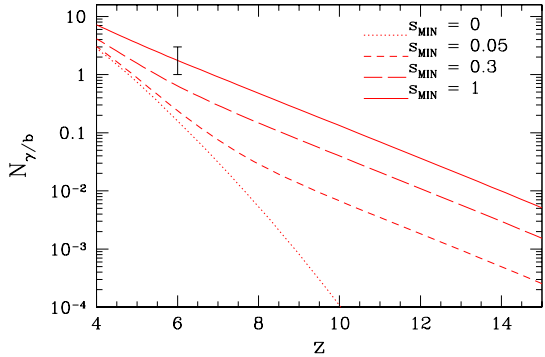}}
\caption{The total ionizing photon-to-baryon ratio as a function
  of redshift for the WMAP3 cosmology for different assumed values of
  the parameter $\smin$: 0 (dotted line), 0.05 (short-dashed
  line, the fiducial value from \citet{gkc07} simulations), 0.3
  (long-dashed line) and 1 (solid line).
} 
\label{figMIN}
\end{figure}

This, somewhat unexpected, result crucially depends on the main
conclusion of \citet{gkc07} that the escape fraction is very small for
low mass galaxies. That conclusion is consistent with the
observational measurements of the escape fraction by \citet{flc03} and
our knowledge of dwarf galaxies in the local universe, which are known
to have large gas fractions and \HI\ extend that exceeds the extend of
the stellar disk. On the other hand, as Figure \ref{figMIN} shows, if
the escape fraction is independent of the galaxy mass ($smin=1$),  the
WMAP3 cosmology comfortably falls into the  
reionization requirement with $N_{\gamma/b}(z=6)\approx1.5$. 

It is, therefore, imperative to have the measurements of the escape fraction
extended to even fainter galaxies and result of
\citet{gkc07} verified with higher resolution simulations and
different numerical methods. Were it found that
the escape fractions of dwarf galaxies are, indeed, negligibly small,
then new, more exotic sources of ionizing radiation (Pop III stars,
X-ray binaries, a new, previously unknown population of faint quasars,
etc\footnote{But {\it not\/} the top-heavy IMF, since the conclusion
  presented in this paper is independent of the stellar IMF and does
  not require any assumption about a particular shape for the IMF.})
would need to be invoked to explain the (relatively) early 
reionization of the universe at $z\approx 6$.

\acknowledgements 

I thank Hsiao-Wen Chen, Andrey Kravtsov, and Jordi Miralda for
valuable comments and corrections to the original manuscript. I am
also grateful to Andrey Kravtsov for the permission to use his halo
mass function code free of charge.
This work was supported in part by the DOE, by the NSF grant
AST-0507596, and by the Kavli Institute for Cosmological Physics at
the University of Chicago.

\bibliographystyle{apj}
\bibliography{gnedin,hizgal,igm,qlf,misc,cosmo}

\end{document}